\documentclass[aps,prl,twocolumn,superscriptaddress,showpacs,floatfix,amsmath,amssymb]{revtex4-1}
\usepackage{graphicx} 
\usepackage{bm}% bold math    
\usepackage{color,mathtools,bbm}    
\usepackage{hyperref}    
\usepackage{float} 
\usepackage{wrapfig}   
\usepackage{mciteplus}
\def\etal{{\em{et al}}}

\DeclareRobustCommand{\specialp}{\ensuremath{\mathcal{P}}}    

\def\F{\mathbf{F}}

\begin{document}

\title{Optical absorption in disordered monolayer molybdenum disulfide}

\author{C. E. Ekuma}
\altaffiliation{National Research Council Research Associate}
\email{cekuma1@gmail.com}
\affiliation{U.S. Naval Research Laboratory, Washington, District of Columbia 20375, USA}

\author{D. Gunlycke}
\email{daniel.gunlycke@nrl.navy.mil}
\affiliation{U.S. Naval Research Laboratory, Washington, District of Columbia 20375, USA}

\date{\today}

\begin{abstract}  
We explore the combined impact of sulfur vacancies and electronic interactions on the optical properties of monolayer MoS$_2$.  First, we present a generalized Anderson-Hubbard Hamiltonian that accounts for both randomly distributed sulfur vacancies and the presence of dielectric screening within the material.  Second, we parameterize this energy-dependent Hamiltonian from first-principles calculations based on density functional theory and the Green function and screened Coulomb (GW) method.  Third, we apply a first-principles-based many-body typical medium method to determine the single-particle electronic structure. Fourth, we solve the Bethe-Salpeter equation to obtain the charge susceptibility $\chi$ with its imaginary part being related to the absorbance $\mathcal{A}$. Our results show that an increased vacancy concentration leads to decreased absorption both in the band continuum and from exciton states within the band gap. We also observe increased absorption below the band gap threshold and present an expression, which describes Lifshitz tails, in excellent qualitative agreement with our numerical calculations.  This latter increased absorption in the $1.0$--$2.5$\,eV makes defect engineering of potential interest for solar cell applications.
\end{abstract}

%, in agreement with recent experiments.
%============================================================================
    
\maketitle
\begin{figure}
    \centering
    \includegraphics[trim = 0mm 0mm 0mm 0mm,scale=0.25,keepaspectratio,clip=true]{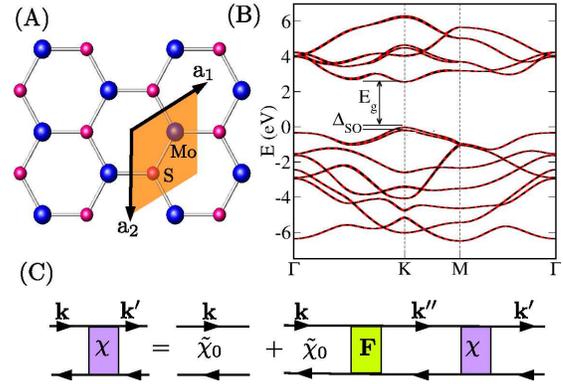}
    \caption{(A) Top view of monolayer MoS$_2$ with the primitive cell defined by the two lattice vectors $\vec{a}_1$ and $\vec{a}_2$ containing 1 Mo and 2 (vertically stacked) S atoms.  (B) DFT+sqGW+SO band structure (dashed black), which exhibits the characteristic direct band gap $E_g=2.6$\,eV and valance-band split $\Delta_\mathrm{SO}=0.22$\,eV at the first Brillouin zone corners (K), is reproduced by the (solid red) bands from the Hamiltonian $\hat H_0$ in Eq.\,(\ref{e.1}).  (C) Diagrammatic representation of the Bethe-Salpeter equation for the TMDCA method, which allows the full, dynamical charge susceptibility $\chi$ to be solved from the screened charge susceptibility $\tilde \chi_0$ and the lattice full vertex function $\F$.}
    \label{f.1}
\end{figure}

 
%\section{Introduction}
Monolayer molybdenum disulfide (MoS$_2$) and other two-dimensional materials are increasingly being explored for modern and future device applications.  This interest originates in part from the atomically thin planar geometry [Fig.\,\ref{f.1}(A)], which allows for the application of well-developed microelectronic, optical, and surface science probing techniques.  Moreover, these materials exhibit a set of attractive electronic properties, which in the case of monolayer MoS$_2$ include a direct band gap in the visible spectrum [Fig.\,\ref{f.1}(B)]. Areas of interest include flexible optoelectronic~\cite{137121,doi:10.1021/nl4007479}, thermoelectric~\cite{doi:10.1021/nl303321g}, valleytronic~\cite{Zeng2012,*Cao2012,Kinetal2012}, and exciton-based applications~\cite{Frindt1963,*PhysRevLett.81.2312,PhysRevLett.111.216805}, as well as solar cells~\cite{Dashora2013216}, photodetectors and light emitters~\cite{Lopez-Sanchez2013,*doi:10.1021/nl400516a,*doi:10.1021/nl301485q}, energy-efficient field-effect transistors (FETs)~\cite{Sunkooketal2012}, and next-generation nanoelectronics~\cite{Radisavljevicetal2011,Lopez-Sanchez2013,Akinwande2014}.
The presence of defects can have a large impact on device performance, both positive and negative. Furthermore, their impact is often more pronounced in low-dimensional materials and in nanoscale devices, which offer far fewer electronic paths around these defects.  While defects can take many forms~\cite{Mak2013,doi:10.1021/acsnano.5b02144,PhysRevB.79.115409,PhysRevB.85.205302,*doi:10.1021/jp075424v}, sulfur vacancies in particular have been observed in large quantities in monolayer MoS$_2$, both in samples produced by micromechanical exfoliation and chemical vapor deposition~\cite{Radisavljevic2013,Hong2015,Chen2015}.  These sulfur vacancies are believed to be the main reason measured carrier mobilities in monolayer MoS$_2$ have not been as high as expected~\cite{Radisavljevicetal2011,nl2043612,doi:10.1021/nl4007479,nl3040042}.

Despite the importance of defects, almost all theoretical studies of monolayer MoS$_2$ and other transition-metal dichalcogenides have so far focused on pristine crystals.  Because of these studies, we understand the main features of the electronic band structures~\cite{PhysRevB.88.245309,*PhysRevB.88.085433,*PhysRevX.4.031005} and the optical absorption spectra~\cite{PhysRevB.79.115409, doi:10.1021/nl903868w,PhysRevLett.105.136805,PhysRevLett.112.216804,*PhysRevB.91.075310,*PhysRevB.88.075434,Frindt1963,*PhysRevLett.81.2312,PhysRevLett.111.216805}.  The effect of many-body interactions on line shifts, excited bound states, carrier dynamics, and optical gain~\cite{C4NR03703K,*ADMA201301244,*acsnano6b03237,*PhysRevB.92.125417} is also under active investigation.  To support the development of any practical applications, however, we also need to develop a better understanding of the properties in disordered MoS$_2$ containing randomly distributed defects.

In this letter, we show that randomly distributed sulfur vacancies could impact the optical properties in several ways.  The presence of vacancies reduces the band-to-band absorption.  It also suppresses the principal exciton peaks, which are delocalized across the monolayers.  More importantly perhaps, we also observe an increased absorption below the band gap, which could be advantageous for solar cell applications that benefit from a large adsorption in the $1.0$--$2.5$\,eV range in the visible and near infrared that dominates the solar irradiance spectrum.

A significant challenge in our investigation is that electron-electron interactions are central to the understanding of many optical features, including the characteristic A and B absorption peaks associated with the generation of excitons. To properly describe these exciton structures, we have extended our recently developed first-principles-based typical medium dynamical cluster approximation (TMDCA) method~\cite{PhysRevLett.118.106404} by explicitly incorporating into the Hamiltonian an energy-dependent and material-specific screened Coulomb interaction $W(E)$ obtained from a separate self-consistent, quasiparticle GW calculations. Furthermore, we have solved the Bethe-Salpeter equation (BSE) [Fig.\,\ref{f.1}(C)] in the particle-hole channel to obtain the charge susceptibility $\chi$ from our self-consistent TMDCA calculations, which account for both the sulfur vacancies and the electron-electron interactions.  This charge susceptibility is a central component in a range of experimental quantities, including optical conductivity, reflectivity measurements, photoluminescence, inelastic neutron (or X-ray) scattering, and electron energy loss spectroscopy.  The imaginary part of the susceptibility is directly related to the absorbance.
%============================================================================
    
%\section{Method}

To capture the physics of defects and electron-electron interactions, we introduce the following energy-dependent Hamiltonian
\begin{equation}
    \hat H(E) = \hat H_0+\sum_{i\alpha\sigma} V_{i\sigma}^\alpha \hat n_{i\sigma}^\alpha + W(E)\sum_{i\alpha} \hat n^\alpha_{i \uparrow} \hat n^\alpha_{i \downarrow},
    \label{e.1}
\end{equation}
where $\hat H_0$ is a Hamiltonian of the pristine crystal, $V_{i\sigma}^\alpha$ is a disorder potential with $i$, $\alpha$, and $\sigma$ being site, orbital, and spin indices, respectively, $\hat n_{i\sigma}^\alpha$ is the number operator, and $W(E)$ is an energy-dependent, crystal-specific screened Coulomb interaction.  This energy-dependent Hamiltonian can be viewed as a generalization of the Anderson-Hubbard Hamiltonian, in which we have substituted $W(E)$ for the energy-independent Hubbard-interaction parameter.

The Hamiltonian for the pristine crystal $\hat H_0$ and the screened Coulomb interaction $W(E)$ can be obtained from calculations based on density functional theory (DFT)~\cite{Hohenberg1964,*Kohn1965} and the GW method~\cite{PhysRevLett.96.226402}.  To simulate the monolayer MoS$_2$ considered herein, we carried out structural relaxation and electronic structure calculations using the {\it Vienna Ab Initio Simulation Package} (VASP)~\cite{KRESSE199615} of periodically repeated monolayers of MoS$_2$ separated by 20\,{\AA}.  These calculations included spin-orbit (SO) interactions and were performed using the Perdew-Burke-Ernzerhof~\cite{PhysRevLett.77.3865} exchange-correlation functional as implemented in VASP. The energy cutoff was 450\,eV and a $15\times 15\times 1$ uniform $\Gamma$-centered grid was used to represent the reciprocal space. Subsequently, we performed self-consistent quasiparticle Green function and screened Coulomb interaction (sqGW) calculations, which also incorporated local field effects beyond the random-phase-approximation using an energy cutoff of 160\,eV. From these calculations, we extracted the screened Coulomb potential $W(E)$ in Eq.\,(\ref{e.1}). Also, the obtained quasiparticle wavefunctions were applied in WANNIER90~\cite{Mostofi2008685} to obtain $\hat H_0$ via a downfolding method that transforms the basis to a set of maximally localized Wannier functions. For monolayer MoS$_2$, we focus on Mo $d$-orbitals and S $p$-orbitals.

There is an excellent agreement between our Hamiltonian $\hat H_0$ and our first-principles calculations, from which it was generated.  The 22 bands produced by the 10 $d$-orbitals on the Mo atom and the 6 $p$-orbitals on each of the two S atoms accurately reproduce the DFT+sqGW+SO band structure around the Fermi level, as shown in Fig.\,\ref{f.1}(B).  This reproduction includes the direct band gap $E_g=2.6$\,eV at the first Brillouin zone corners (K) and the valence band split $\Delta_\mathrm{SO}=0.22$\,eV.

The randomly distributed sulfur vacancies in the system are modeled through the disorder potential $V_{i\sigma}^\alpha$ in Eq.\,(\ref{e.1}).  Assuming no explicit orbital and spin dependence, the disorder potential $V_{i\sigma}^\alpha$ reduces to a site potential $V_i$.  We generated this site potential from a two-point distribution with $V_i\in\{0,V_\mathrm{vac}\}$, where the two elements are the potentials on non-vacant and vacant sites, respectively.  To prevent any electron occupation on vacant sites, we set the potential $V_\mathrm{vac}$ to be far greater than the material bandwidth.  We further, let the distribution to be unbalanced with the probability that an arbitrary site is vacant being equal to the sulfur vacancy concentration parameter $\delta$.

To determine the single-particle properties of the many-body Hamiltonian above, we adopt an approach based on the dynamical mean-field theory. This theory maps the disordered lattice onto an effective medium characterized by a hybridization function obtained through a set of self-consistency equations.  Herein, we solve the self-consistency equations within the typical medium dynamical cluster approximation (see Refs.~\cite{PhysRevB.89.081107,*PhysRevB.92.014209,*0953-8984-26-27-274209,*PhysRevB.92.205111,*PhysRevB.94.224208,EkumaDissertation2015} for additional information) to obtain the electron-interaction and sulfur-vacancy-dressed retarded single-particle Green function, following the approach described in Ref.~\onlinecite{PhysRevLett.118.106404}. The advantage of using the TMDCA method in obtaining the single-particle Green function is that it is based on geometric averages, which unlike the more common arithmetic averages, allows for the discrimination of local and extended states~\cite{PhysRevLett.118.106404,PhysRevB.89.081107,*PhysRevB.92.014209,*0953-8984-26-27-274209,*PhysRevB.92.205111,*PhysRevB.94.224208,PhysRevB.92.201114,Vlad2003}.  This discrimination is critical for phenomena involving electron localization.

Obtaining full convergence of exciton features~\cite{PhysRev.52.191,*Mott1938,PhysRev.37.17} in monolayer MoS$_2$ is challenging. It is well-known that an unusually fine reciprocal space grid is required~\cite{PhysRevLett.111.216805}. Owing to the resulting computational demand, we have not been able to increase the cluster size within the TMDCA method sufficiently to reach simultaneous convergence of the reciprocal grid and cluster size. Therefore, to avoid the risk of potentially spurious spatial correlation effects interfering with the exciton features, we present herein results obtained using single-site clusters and a hexagonal grid consisting of $50\times 50\times 1$ $\vec k$ points.

The TMDCA self-consistency equations produce the single-particle retarded Green functions in the presence of both electronic interactions and sulfur vacancies, which we used as input in our two-particle calculations. The procedures to obtaining the full lattice-dynamical susceptibility goes as follows: (i) Using the single-particle retarded Green function $G(\vec k,E)$ calculated self-consistently from the TMDCA equations, we obtain the spectral function $A(\vec k,E) = -\frac{1}{\pi} \mathfrak{Im}\,G(\vec k,E)$, which we subsequently use to calculate the imaginary part of the bare dynamic charge susceptibility $\chi_0(\vec q,\omega)$. For electron-hole excitations, we have
\begin{widetext}
\begin{align}
	\mathfrak{Re}\chi_0(\vec q,\omega) &= \frac{1}{\pi}\specialp \int_{-\infty}^{+\infty}\frac{\mathfrak{Im}\,\chi_0(\vec q,\omega)\mathrm{d}\omega'}{\omega'-\omega},\nonumber\\
	\mathfrak{Im}\,\chi_0(\vec q,\omega) &= -2\pi \sum_{\vec k} \int_{-\infty}^{+\infty} [f(E)-f(E+\hbar\omega)]\,A(\vec k+\vec q,E+\hbar\omega)\,A(\vec k,E)\,\mathrm{d}E,
	\label{e.2}
\end{align}
\end{widetext}
where $\specialp$ denotes Cauchy principal value and $f(E)$ is the Fermi function. Since our lattice problem is solved in real space and for the fact that it is numerically more advantageous to work in real frequency, we have obtained Eq.~\ref{e.2} using the spectral representation~\cite{PhysRevB.92.201114}. We note that Eq.~\ref{e.2} is the real space equivalence of the usual Matsubara frequency form that could be depicted as $\chi_0(\vec q, i\omega) = \frac{T}{N} \sum_{\vec k, i E} \,G(\vec k+\vec q,iE+ i \hbar\omega)\,G(\vec k,iE)$~\cite{Negele,*fetter1971quantum}. (ii) To account for the screening caused by the electric polarization that results from the excitations inside the material, the bare susceptibility has been renormalized as $\tilde \chi_0(\vec q,\omega)=\chi_0(\vec q,\omega)\left[\mathbbm{1}-W(\omega)\chi_0(\vec q,\omega)\right]^{-1}$, where $\mathbbm{1}$ is the identity matrix. (iii) To obtain the full lattice-dynamical susceptibility, we further need the full lattice vertex function $\F(\vec q,\omega) = \Gamma(\vec q, \omega) [\mathbbm{1} -  \tilde \chi_0(\vec q,\omega)\Gamma(\vec q,\omega)]^{-1}$, where $\Gamma(\vec q,\omega)$ is the irreducible lattice vertex function, which we approximate using the one obtained within the TMDCA method~\cite{Jarrell01,*PhysRevB.61.12739,*RevModPhys.77.1027,*PhysRevLett.69.168,*zhang2017,EkumaDissertation2015}. (iv) The vertex function $\F(\vec q,\omega)$ is used along with the renormalized $\tilde \chi_0(\vec q,\omega)$ in the BSE depicted in Fig.\,\ref{f.1}(C):
\begin{align}
	\langle \vec k|\chi|\vec k'\rangle &= \langle \vec k|\tilde \chi_0|\vec k\rangle+\sum_{\vec k''}\langle \vec k|\tilde \chi_0|\vec k\rangle\langle \vec k|\F|\vec k''\rangle\langle \vec k''|\chi|\vec k'\rangle.
	\label{e.3}
\end{align}
(v) Solving Eq.~\ref{e.3} yields the full lattice-dynamical charge susceptibility $\chi(\vec q,\omega)$. Below, we focus on the momentum integrated full charge susceptibility $\chi(\omega)$.

%============================================================================

%\section{Results}

\begin{figure}[h!]
    \includegraphics[trim = 0mm 0mm 0mm 0mm,scale=0.5,keepaspectratio,clip=true]{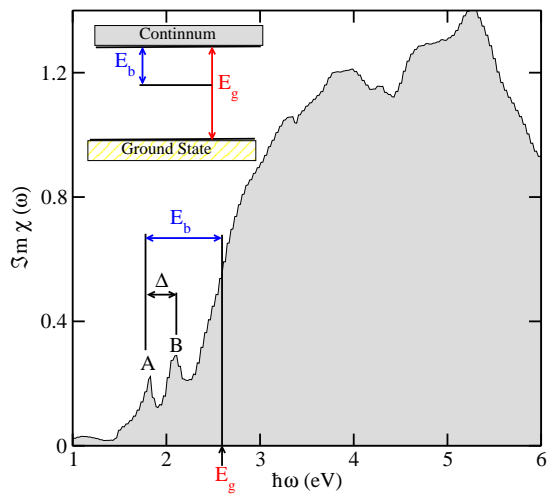}
    \caption{Imaginary part of the charge susceptibility for pristine monolayer MoS$_2$ as a function of the excitation energy $\hbar\omega$.  A and B denote the main two exciton peaks, separated by $\Delta=0.24$\,eV and found below the band gap energy $E_g=2.6$\,eV.  $E_b=0.8$\,eV is the binding energy of A exciton.}
    \label{f.2}
\end{figure}

We show in Fig.~\ref{f.2} the imaginary part of the charge susceptibility $\mathfrak{Im}\,\chi (\omega)$ for the pristine monolayer MoS$_2$. In addition to the large contribution for excitation energies above the band gap associated with the generation of free electron/hole pairs, this susceptibility exhibits the two dominant exciton peaks observed in MoS$_2$, A and B (our resolution is not sufficient for the observation of trion and excited exciton features).  The separation between A and B peaks is $\Delta=0.24$\,eV, which is only slightly larger than the calculated valence-band split of $\Delta_\mathrm{SO}=0.22$\,eV, mentioned above.  While the position of the A exciton peak $E_\mathrm{A}=1.84$\,eV is in excellent agreement with experiment, we caution readers not to over-interpret its significance as the agreement is likely somewhat coincidental. Assuming the band gap $E_g=2.6$\,eV from our DFT$+$sQGW$+$SO calculations, the position of the A peak $E_\mathrm{A}=1.84$\,eV corresponds to the A exciton binding energy $E_b = 0.66$\,eV.  This value is the upper limit of experimental values, which are in the range of 0.2--0.7 eV~\cite{Radisavljevic2013,PhysRevB.94.075440,*10.1021/nl504868p,*10.1021/nl501133c,*Klots2014,*Yu2015}, presumably because our calculations do not consider screening from the substrate.

\begin{figure}
    \includegraphics[trim = 0mm 0mm 0mm 0mm,scale=0.5,keepaspectratio,clip=true]{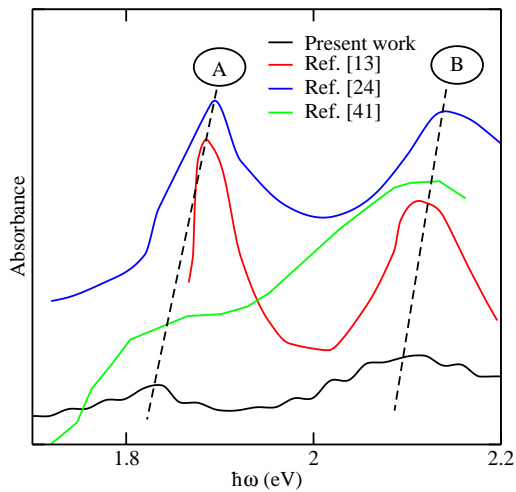}
    \caption{ The absorbance $\mathcal{A}(\omega)$ obtained using Eq.~\ref{e.4} for pristine MoS$_2$ monolayer as compared with experimental data. Observe that the relative trend and position of the principal exciton peaks are nicely reproduced in our computed data.}
    \label{f.2a}
\end{figure}

We can directly relate our calculations to experiment by calculating the thickness-independent absorbance $\mathcal{A}(\omega)$, using
\begin{equation}
	\mathcal{A}(\omega)=\frac{\omega\ell}{c}\,\mathfrak{Im}\,\chi(\omega),
	\label{e.4}
\end{equation}
where $\omega$ is the angular frequency of the excitation, $c$ is the speed of light, and $\ell$ is an arbitrary vertical length unit that factor out, as the susceptibility $\chi\propto \ell^{-1}$~\cite{PhysRevB.87.035438}.  Figure~\ref{f.2a} shows good qualitative agreement between our calculated spectrum and typical experimental spectra~\cite{Mak2013,PhysRevLett.105.136805,1347-4065-55-3-038003}.  Small differences in peak positions and heights are common and to be expected.  These could originate from different conditions on the experimental side and different assumptions on the theoretical side.

\begin{figure}[b!]
    \includegraphics[trim = 0mm 0mm 0mm 0mm,scale=0.5,keepaspectratio,clip=true]{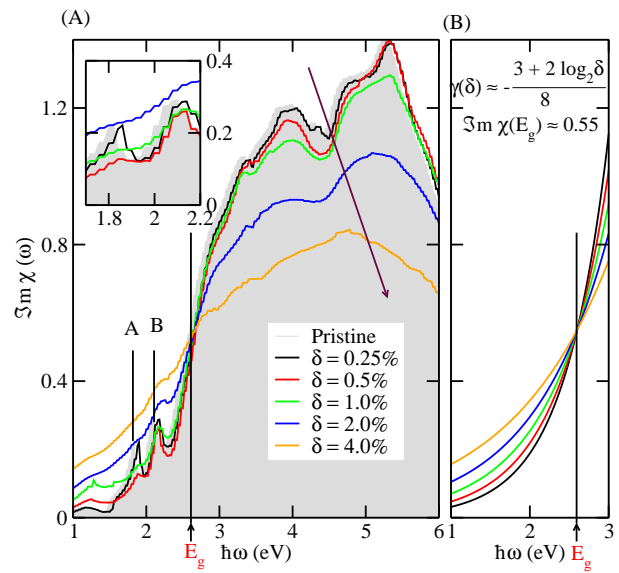}
    \caption{Imaginary part of the charge susceptibility for monolayer MoS$_2$ at various sulfur vacancy concentrations $\delta$ from (A) our calculations and (B) Eq.\,\ref{e.5}.  Increasing the concentration leads to a deterioration of the A and B exciton peaks (see inset) and suppresses the generation of free electron-hole pairs (see arrow). On the other hand, increasing the amount of sulfur vacancies generally increases $\mathfrak{Im}\,\chi (\omega)$, and hence the optical absorption below the band gap $E_g=2.6$\,eV.}
    \label{f.3}
\end{figure}

To explore the role of randomly distributed sulfur vacancies on the optical absorption spectrum, we show in Fig.~\ref{f.3} the same imaginary part of the charge susceptibility as in Fig.~\ref{f.2} but for various sulfur vacancy concentrations $\delta$.  Again, we see the peak structure from the A and B excitons. The peaks, however, gradually lose intensity as $\delta$ increases and eventually vanish at approximately $\delta=4.0$\%. The disappearance of the principal exciton peaks at high disorder has recently been observed in experiment; e.g., Kang \etal~\cite{10.1021/jp506964m} reported the disappearance of the principal exciton peaks at high plasma exposure times in their oxygen plasma treatment of monolayer MoS$_2$. Some experiments have reported that extrinsic defects induced by e.g., electron-irradiation or ion bombardment could diminish the exciton peaks~\cite{10.1021/acsnano.5b07388,*PhysRevB.91.195411,*C4NR02142H}. Several other experiments, e.g., Refs.~\cite{Mak2013,1347-4065-55-3-038003,10.1021/nl4011172,vanderZande2013} reported significant decrease in the intensity of the exciton peaks due to unintentional defect states in their samples.

There also appears to be a slight blue shift of the exciton peaks in Fig.~\ref{f.3}. This is also consistent with some experimental studies that have reported blue shift of the A-peak in monolayer MoS$_2$ that is mainly due to defects~\cite{10.1021/acs.jpcc.6b06828}. 

The impact of the sulfur vacancies on the overall magnitude of $\mathfrak{Im}\,\chi (\omega)$ varies in different parts of the spectrum.  Above the band gap transition energy threshold, Fig.~\ref{f.3} shows a general decrease in $\mathfrak{Im}\,\chi (\omega)$ as the vacancy concentration increases. This is expected as the disorder introduced by the vacancies breaks up the extended states within the conduction and valence bands, leading to less generation of free electron-hole pairs.

We find that the susceptibility below the band gap threshold, in contrast to that above, generally increases with an increased vacancy concentration. This is a manifestation of an increased number of electronic states within the band gap, especially near the band edges. This effect is well-known in disordered systems, where the characteristic exponential dependence on energy is known as Lifshitz tails~\cite{10.1080/00018736400101061}. In our system, such states are caused by local perturbations of the electron-electron interactions around the vacancies. For a sufficiently large vacancy concentration, we expect the susceptibility in this sub-threshold regime to be dominated by excitations between such states. To lowest order, we would then have
\begin{equation}
	\mathfrak{Im}\,\chi(\omega)\approx\mathfrak{Im}\,\chi(E_g)\,\mathrm{e}^{\gamma(\delta)[\hbar\omega-E_g]},
	\label{e.5}
\end{equation}
for a two-dimensional system such as monolayer MoS$_2$.  We used $\mathfrak{Im}\,\chi(E_g)\approx 0.55$ and $\gamma(\delta)\approx-(1/8)(3+2\log_2\delta)\,\mathrm{eV}^{-1}$ in Fig.~\ref{f.3} and obtained an excellent qualitative agreement with our first-principles-based calculations.

In summary, we have studied the role of randomly distributed sulfur vacancy and material-specific Coulomb interactions on the optical properties of monolayer MoS$_2$ using a first-principles-based many-body typical medium approach. Our results show that both electronic interactions and sulfur vacancies affect the dynamic charge susceptibility, and concomitantly the optical absorption. In particular, we observe an increased susceptibility below the band gap threshold with an increased vacancy concentration. This finding suggests that defect engineering could be useful for optoelectronic applications, including solar cell applications.

%Second, electrically controlled absorption of light and photoluminescence in high-mobility MoS$_2$ can be utilized to create nanoscale optoelectronic modulators operating in the visible range. 
%Concurrently, we observe a decrease in absorption at the same wavelength. We propose that these phenomena are caused by the interaction of excitons in MoS2 with conduction electrons via the phase-space filling effect.
%\begin{acknowledgments}
This work has been funded by the Office of Naval Research, directly and through the U.S. Naval Research Laboratory.  CEE acknowledges support from the NRC Research Associateship Programs.  \\ 
%\end{acknowledgments}

\end{document}